\documentclass[12pt,amsbsy]{article}
\newcommand{\beq}{\begin{equation}}
\newcommand{\eeq}{\end{equation}}
\newcommand{\beqn}{\begin{eqnarray}}
\newcommand{\eeqn}{\end{eqnarray}}

\newcommand{\al}{\mbox{${\alpha}$}}

\newcommand{\ga}{\mbox{${\gamma}$}}

\newcommand{\de}{\mbox{${\delta}$}}

\newcommand{\om}{\mbox{${\omega}$}}

\newcommand{\na}{\mbox{${\nabla}$}}
\newcommand{\pa}{\mbox{${\partial}$}}

\begin{document}

\begin{center}
{\bf \large Remark on Immirzi Parameter, Torsion, and Discrete
Symmetries}
\end{center}

\begin{center}
I.B. Khriplovich\footnote{khriplovich@inp.nsk.su} and A.A.
Pomeransky\footnote{a.a.pomeransky@inp.nsk.su}
\end{center}
\begin{center}
Budker Institute of Nuclear Physics\\
630090 Novosibirsk, Russia,\\
and Novosibirsk University
\end{center}

\bigskip

\begin{abstract}
We point out that the new interaction of spinning particles with
the torsion tensor, discussed recently, is odd under charge
conjugation and time reversal. This explains rather unexpected
symmetry properties of the induced effective 4-fermion
interaction.
\end{abstract}

\vspace{8mm}

As has been demonstrated in \cite{hol},  the gravitational action
can be written as
\beq\label{sg}
S_g = -\,\frac{1}{16\pi G}\,\int d^4 x \,e\, e^{\mu}_I\,
e^{\nu}_J\left( R^{IJ}_{\mu\nu}\,
-\,\frac{1}{\ga}\,^{*}R^{IJ}_{\mu\nu}\right).
\eeq
Here $I,J=0,1,2,3$ are internal Lorentz indices, $\mu,\nu=0,1,2,3$
are space-time indices; $e_\mu^I$ is the tetrad field, $e$ is its
determinant, and $e^{\mu}_I$ is the object inverse to $e_\mu^I$;
the curvature tensor is
\[
R^{IJ}_{\mu\nu}=-\,\pa_{\mu} \om^{IJ}_{\;\;\;\;\nu}+\pa_{\nu}
\om^{IJ}_{\;\;\;\;\mu}
+\om^{IK}_{\;\;\;\;\mu}\om^{\;\;\;J}_{K\;\;\nu}-\om^{IK}_{\;\;\;\;\nu}\om^{\;\;\;J}_{K\;\;\mu}\,,
\]
with its dual one
\[
^{*}R^{IJ}_{\mu\nu}=
\,\frac{1}{2}\,\varepsilon^{IJ}_{KL}R^{KL}_{\mu\nu}\,;
\]
$\om^{IJ}_{\mu}$ is the connection. As to the so-called Immirzi
parameter $\ga$, the common belief is that it does not enter
equations of motion since it appears as a factor at the term that
vanishes on mass shell.

However, it was pointed out recently in \cite{pero} (see also
earlier paper \cite{hojman}) that in the presence of spin 1/2
particles the second term in action (1) does not vanish on mass
shell, and therefore $\ga$ should enter the corresponding
equations of motion. 

Then it was pointed out in
\cite{fmt} (see also \cite{sha}) that the fermion action can be generalized as
follows:
 \beq\label{sf} S_f = \int d^4 x \,e\, \frac{1}{2}\left[
 (1-i\al)\,\bar{\psi}\,\ga^I\,e^{\mu}_I\,i\na_\mu \psi -
 (1+i\al)\,i\, \overline{\na_\mu\psi}\,\ga^{I}e_{I}^\mu\psi\right].
 \eeq
Here $\ga^I$ are the Dirac matrices;
 \[
 \na_\mu = \pa_\mu -
 \frac{1}{4}\,\om^{IJ}_{\;\;\;\;\mu}\,\ga_I\ga_J\,,\quad
 [\na_\mu,\na_\nu] = \frac{1}{4}\,R^{IJ}_{\mu\nu}\,\ga_I\ga_J.
 \]
As to the real constant
$\al$, it is of no consequence, generating a total derivative
only, if the theory is torsion free, i.e. if
 \beq\label{tf}
 \na_{[\mu}e_{\nu]}^I =0.
 \eeq
It is pointed out in \cite{fmt,sha}, however, that $\al$ becomes
operative in the presence of the so-called contorsion tensor
$C^{IJ}_{\;\;\;\;\mu}$ contributing to $\om^{IJ}_{\;\;\;\;\mu}$,
so that instead of (\ref{tf}) we have
 \beq\label{cont}
 \na_{[\mu}e_{\nu]}^I = -\,C^{IJ}_{\;\;\;\;[\mu}\,e_{\nu]J}\,.
 \eeq
In this case the structure generated by $\al$ does not reduce to
the total derivative:
 \beq
 e\,(\bar{\psi}\,\ga^I\,e^{\mu}_I\,\na_\mu \psi +
 \overline{\na_\mu\psi}\,\ga^{I}e_I^\mu\psi) = \pa_\mu
 (e\,e^{\mu}_I\,V^I) - e e^{\mu}_I\,C^I_{\;J\mu}V^J,
 \eeq
where $V^J$ is the vector current,
 $V^J=\bar{\psi}\,\ga^J\,\psi\,$.
Thus, nonvanishing $\al$ generates the effective interaction of
spinning particles
 \beq\label{sal} S_{\alpha} =-\, \frac{\al}{2}\,
 \int d^4 x\,e\, e^{\mu}_I\,C^I_{\;J\mu}V^J =-\, \frac{\al}{2}\,\int
 d^4 x\,e\, e^{\mu}_I\,C^I_{\;J\mu}\,\bar{\psi}\,\ga^J\,\psi\,,
 \eeq
new as compared to the canonical one.

Finally, the effective 4-fermion interaction obtained in this way
in \cite{fmt} is
\beq\label{ff}
S_{ff}= \,\frac{3}{2}\,\pi G\,\frac{\ga^2}{\ga^2 + 1}\,\int
d^4x\,e \left(A^2 +\,\frac{2\,\al}{\ga}VA -\al^2 V^2\right),
\eeq
where $A$ is the fermion axial current, $A^I =
\bar{\psi}\,\ga^5\,\ga^I\,\psi\,$. For $\al=0$ and $\ga \to
\infty$, the result (\ref{ff}) goes over into that derived in
\cite{weyl,ki}.

In connection with equation (\ref{ff}), the following question
arises. The usual curvature tensor $R^{IJ}_{\mu\nu}$ and its dual
$^*R^{IJ}_{\mu\nu}$ behave in opposite ways both under the space
reflection $P$ and the time reversal $T$ (see, e.g.,
\cite{hojman}). To see it, we recall firstly that the duality
transformation in electrodynamics results in ${\bf E} \to {\bf
B}$, ${\bf B} \to -{\bf E}$, and that ${\bf E}$ and ${\bf B}$
behave in opposite ways both under $P$ and $T$. Then our duality
problem in gravity is effectively reduced to that in
electrodynamics by using the Petrov decomposition for the
curvature tensor. Now, how it comes that the $T$-odd (and
therefore $CP$-odd, in virtue of the $CPT$ theorem), $P$-odd
second term in (1) generates the $P$-odd, but $T$-even $VA$
interaction in (\ref{ff})?

The answer consists in rather special properties of the new
effective interaction (\ref{sal}) introduced in \cite{fmt,sha}. The
factor $e^{\mu}_I\,C^I_{\;J\mu}$ therein is the same for fermions
and antifermions, i.e. it is even under charge conjugation $C$.
Then, the vector current $V^J = \bar{\psi}\,\ga^J\,\psi\,$ is
well-known to be $C$-odd. Thus, action (\ref{sal}) as a whole is
$C$-odd too.

On the other hand, interaction (\ref{sal}) is also $T$-odd.
Indeed, to simplify the arguments, let us consider a specific case
when the covariant vector current $V_J$ has only a single, time
component $\rho$, and $e^{\mu}_I$ reduces to $\de^{\mu}_I$. Then
\[
e^{\mu}_I\,C^I_{\;J\mu}\,V^J\, \to \; C_m^{mt}\,\rho\,, \quad
m=1,2,3\,.
\]
Now, it can be easily demonstrated that the invariance of the
scalar curvature $R$ under time reversal dictates that under this
procedure $\om_m^{mt}$ should change sign. Obviously, $C_m^{mt}$
has the same property. Since the particle density $\rho$ does not
change under time reversal, interaction (\ref{sal}) is $T$-odd
indeed.

Therefore, it is only natural that the $P$-odd and $T$-odd part of
gravitational action (1), being combined with $T$-odd action
(\ref{sal}), generates the $VA$ term in (\ref{ff}).

Our last remark refers to two other terms in the 4-fermion
interaction (\ref{ff}). The coefficients at both are even
functions of $1/\ga$, which means that they effectively arise via
even powers of $P$- and $T$-odd term in (1). Besides, the
coefficient at $V^2$ is proportional to $\al^2$, i.e. is of second
order in the $C$-odd interaction (\ref{sal}). No wonder therefore
that both these terms are invariant under all the three discrete
symmetries, $C$, $P$, and $T$. On the other hand, it is only
natural in this respect that the coefficient at $VA$ is an odd
function of $1/\ga$ and is linear in $\al$.

\begin{center}***\end{center}

The investigation was supported in part by the Russian Foundation
for Basic Research through Grant No. 05-02-16627.

\end{document}